\documentclass[nofootinbib,preprint]{revtex4}%
\usepackage{hyperref}
\usepackage{amsmath}
\usepackage{amsfonts}
\usepackage{amssymb}
\usepackage{graphicx}%
\begin{document}
\title{Semiclassical Analysis of String/Gauge Duality\\on Non-commutative Space}
\author{R. C. Rashkov}
\email{rash@phys.uni-sofia.bg}
\affiliation{Department of Physics, Sofia University, 1164 Sofia, Bulgaria}
\author{K. S. Viswanathan}
\email{kviswana@sfu.ca}
\affiliation{Department of Physics, Simon Fraser University, Burnaby BC, Canada}
\author{Yi Yang}
\email{yiyang@mail.nctu.edu.tw}
\affiliation{Department of Electrophysics, National Chiao Tung University, Hsinchu, Taiwan.}
\date{\today }

\begin{abstract}
We use semiclassical methods to study closed strings in the modified
$AdS_{5}\times S^{5}$ background with constant $B$ fields. The point-like
closed strings and the stretched closed strings rotating around the big circle
of $S^{5}$ are considered. Quantization of these closed string leads to a
time-dependent string spectrum, which we argue corresponds to the RG-flow of
the dual noncommutative Yang-Mills theory.

\end{abstract}
\maketitle

\section{Introduction}

AdS/CFT duality \cite{9711200} is a powerful method to study field theories by
studying their dual string theories. Usually, full string theories in the
curved background, such as $AdS_{5}\times S^{5}$, are not solvable. We have to
study their low energy effective theories, i.e. supergravity theories.
According to IR/UV duality, the low energy region in the string theories is
dual to the high energy region in the corresponding field theories side.
However, the low energy region of the field theory is more attractive for
practical reason. To study the low energy region of these field theories, we
need to solve the full string theory. It was shown that by taking Penrose
limit, $AdS_{5}\times S^{5}$ geometry reduces to a pp-wave geometry
\cite{0202021}, which is solvable for the full string theory.

The success of pp-wave method for $AdS_{5}\times S^{5}$ background led many
authors to use it to study some more complicated geometry backgrounds e.g.
\cite{0308024}. For example, the near horizon geometry of NS5-branes is the
linear dilaton background. String theory in this background was conjectured to
be dual to a non-local field theory \cite{9705221}, so called Little String
Theory (LST). It was shown that the pp-wave limit of this background is the
Nappi-Witten background. String theory in Nappi-Witten background is again
exactly solvable and has been widely studied, see for example
\cite{0205258,0305028}. A more complicated example is the pp-wave limit of the
near horizon geometry of non-extremal NS5-branes, which leads to a time
dependent background. Luckily, strings in this time dependent background is
also exactly solvable and has been studied in \cite{0205258,0211289}. Another
non-local gauge theory example is non-commutative Yang-Mills theories (NCYM),
which are conjectured to be dual to the near horizon geometry of $N$ coincide
D$p$-branes in the presence of the background constant $B$ fields. Geometries
with different number $p$ have been studied in \cite{9910092,0205258} and more
extensively in \cite{0209054}, in which the authors claimed that only for
D6-brane, the pp-wave limit of the geometry leads to a \textit{time
independent} background and is solvable. In this work, we will show that for
general D$p$-branes, pp-wave limit of the geometry leads to time dependent
backgrounds which are solvable in certain limit. Studying string theories in
these backgrounds will enable us to understand more about the dual theory -
noncommutative Yang-Mills theory, such as IR/UV mixing, non-local behavior
etc.. In addition, through the holography relationship, we can use the time
dependent pp-wave background to study the features of RG flow in NCYM.

In the next section, we study the supergravity solution with constant NS $B$
fields. In section 3, semiclassical method is used to study the first order
fluctuation around some classical string configurations. The Hamiltonian will
be obtained for the fluctuation. We then quantize the model to get the string
spectrum in section 4. Section 5 contains discussion on the results.

\section{Supergravity solution with constant $B$ fields}

In this section we consider a system of $Dp$-branes with a constant NS $B$
along their world-volume directions.. For simplicity, we consider the $B$
field only along the directions of $x_{B}^{2}$ and $x_{B}^{3}$ in the
$D$-brane, the supergravity solution can be obtained by performing a T-duality
along $x^{3}$ and then T-dualize back. The solution in string metric is
\cite{9907166,9908134}
\begin{equation}
ds_{B}^{2}=f^{-1/2}\left[  -dx_{0}^{2}+dx_{1}^{2}+h\left(  dx_{B2}^{2}%
+dx_{B3}^{2}\right)  \right]  +f^{1/2}\left(  dr^{2}+r^{2}d\Omega_{5}%
^{2}\right)  ,
\end{equation}
where
\begin{align}
f  &  =1+\dfrac{\alpha^{\prime2}R^{4}}{r^{4}},\text{ \ \ \ \ \ \ \ \ }%
h^{-1}=\sin^{2}\theta f^{-1}+\cos^{2}\theta,\nonumber\\
B_{23}  &  =\dfrac{\sin\theta}{\cos\theta}f^{-1}h,\text{ \ \ \ \ \ \ \ \ \ }%
e^{2\phi}=g^{2}h,\nonumber\\
F_{01r}  &  =\dfrac{1}{g}\sin\theta\partial_{r}f^{-1},\text{ \ \ }%
F_{0123r}=\dfrac{1}{g}\cos\theta h\partial_{r}f^{-1}. \label{transfer1}%
\end{align}
To study the dual noncommutative Yang-Mills theory on the boundary, we need to
take the background $B$ field to infinity $B_{23}\rightarrow\infty$. Following
the procedure in \cite{9907166,9908134}, we first rescale the parameters as
follows
\begin{align}
x_{0,1}  &  \rightarrow x_{0,1},\text{ \ }x_{B2,3}\rightarrow\left(
\dfrac{\alpha^{\prime}}{b}\right)  ^{-1}x_{B2,3},\text{ \ \ }B_{23}%
\rightarrow\left(  \dfrac{\alpha^{\prime}}{b}\right)  ^{2}B_{23}\nonumber\\
\tan\theta &  =\dfrac{b}{\alpha^{\prime}},\text{ \ \ }r=\alpha^{\prime}%
R^{2}u,\text{ \ \ \ \ }g\rightarrow\dfrac{\alpha^{\prime}}{b}g,
\label{transfer2}%
\end{align}
then we take the decoupling limit $\alpha^{\prime}\rightarrow0$ with $b$
fixed. The metric now can be written on as
\begin{equation}
ds_{B}^{2}=\alpha^{\prime}R^{2}\left[  u^{2}\left(  -dx_{0}^{2}+dx_{1}%
^{2}\right)  +\dfrac{u^{2}}{1+a^{4}u^{4}}\left(  dx_{B2}^{2}+dx_{B3}%
^{2}\right)  +\dfrac{du^{2}}{u^{2}}+d\Omega_{5}^{2}\right]  , \label{metric}%
\end{equation}
where%
\begin{align}
a^{2}  &  =bR^{2},\text{ \ \ }B_{23}=\dfrac{\alpha^{\prime}}{b}\dfrac
{a^{4}u^{4}}{1+a^{4}u^{4}},\text{ \ \ }e^{2\phi}=\dfrac{g^{2}}{1+a^{4}u^{4}%
},\nonumber\\
&  A_{01}=\alpha^{\prime}\dfrac{b}{g}u^{4}R^{4},\text{ \ \ }F_{0123u}%
=\dfrac{\alpha^{\prime2}\partial_{u}\left(  u^{4}R^{4}\right)  }{g\left(
1+a^{4}u^{4}\right)  }. \label{transfer3}%
\end{align}
We will consider sigma model in the above background%
\begin{equation}
S=\frac{1}{4\pi\alpha^{\prime}}\int d^{2}\sigma\left(  \sqrt{h}h^{ab}%
G_{mn}\partial_{a}X^{m}\partial_{b}X^{n}+\epsilon^{ab}B_{ij}\partial_{a}%
X_{B}^{i}\partial_{b}X_{B}^{j}\right)  , \label{sigma action}%
\end{equation}
where the metric $G_{mn}$\ and the constant $B_{ij}$ are given by equation
(\ref{metric}) and (\ref{transfer3}) respectively. Let's make some
transformations to bring the metric to a more convenient form. First of all,
we set $R=1$ for simplify and define the variables
\begin{equation}
x_{B}^{i}=\sqrt{1+a^{4}u^{4}}x^{i},i=2,3. \label{xB-x}%
\end{equation}
In these notations the metric can be written as
\begin{equation}
ds_{B}^{2}=ds^{2}+\dfrac{4a^{8}x^{i2}u^{8}}{\left(  1+a^{4}u^{4}\right)  ^{2}%
}du^{2}+\dfrac{4a^{4}u^{5}x^{i}}{1+a^{4}u^{4}}dx^{i}du \label{modified metric}%
\end{equation}
where
\begin{equation}
ds^{2}=u^{2}\left(  -dx_{0}^{2}+dx_{1}^{2}+dx_{2}^{2}+dx_{3}^{2}\right)
+\dfrac{du^{2}}{u^{2}}+d\Omega_{5}^{2} \label{new1}%
\end{equation}
is the standard $AdS_{5}\times S^{5}$ metric. We call the above metric
(\ref{modified metric}) modified $AdS_{5}\times S^{5}$ metric because the only
deviation of \ref{new1} from the standard $AdS_{5}\times S^{5}$ metric is
given by some $B$ field dependent terms. Later we will consider the case of
$a\rightarrow0$, where the modification from the standard $AdS_{5}\times
S^{5}$ metric is small provided $u$ is not too large.

Next, to write the above metric in \textquotedblleft global\textquotedblright%
\ coordinates, we use the transformations
\begin{align}
u  &  =\cosh\rho\cos t-\sinh\rho\Omega_{4},\nonumber\\
x^{0}  &  =\dfrac{\cosh\rho\sin t}{\cosh\rho\cos t-\sinh\rho\Omega_{4}%
},\nonumber\\
x^{\mu}  &  =\dfrac{\sinh\rho\Omega_{\mu}}{\cosh\rho\cos t-\sinh\rho\Omega
_{4}}, \label{globle}%
\end{align}
where the four-vector $\Omega$ satisfies the normalization condition
$\Omega^{2}=1$ and the explicit form of its components is
\begin{align}
\Omega_{1}  &  =\cos\beta_{1}\cos\beta_{2}\cos\beta_{3},\nonumber\\
\Omega_{2}  &  =\cos\beta_{1}\cos\beta_{2}\sin\beta_{3},\nonumber\\
\Omega_{3}  &  =\cos\beta_{1}\sin\beta_{2},\nonumber\\
\Omega_{4}  &  =\sin\beta_{1}.
\end{align}
The metric now takes the form as
\begin{align}
ds_{B}^{2}  &  =-\cosh^{2}\rho dt^{2}+d\rho^{2}+\sinh^{2}\rho d\Omega_{3}%
^{2}\nonumber\\
&  \text{ \ \ }+d\psi_{1}^{2}+\cos^{2}\psi_{1}\left(  d\psi_{2}^{2}+\cos
^{2}\psi_{2}d\Omega_{3}^{\prime2}\right) \nonumber\\
&  \text{ \ \ }+\dfrac{4a^{8}x^{i2}u^{8}}{\left(  1+a^{4}u^{4}\right)  ^{2}%
}du^{2}+\dfrac{4a^{4}u^{5}x^{i}}{1+a^{4}u^{4}}dx^{i}du, \label{globle metric}%
\end{align}
where
\begin{align}
d\Omega_{3}^{2}  &  =d\beta_{1}^{2}+\cos^{2}\beta_{1}\left(  d\beta_{2}%
^{2}+\cos^{2}\beta_{2}d\beta_{3}^{2}\right)  ,\nonumber\\
d\Omega_{3}^{\prime2}  &  =d\psi_{3}^{2}+\cos^{2}\psi_{3}\left(  d\psi_{4}%
^{2}+\cos^{2}\psi_{4}d\psi_{5}^{2}\right)  .
\end{align}
In the final form of the metric (\ref{globle metric}), to keep the metric
neat, we did not write $dx^{i}$ and $du$ in terms of the new coordinates
explicitly in the last two terms. Reader should be careful that there are only
10 independent coordinates in the metric (\ref{globle metric}) since $u$ and
$x^{i}$ $\left(  i=2,3\right)  $ are not independent coordinates.

\section{Semiclassical analysis}

In this section we will study two classical solutions of the sigma model
(\ref{sigma action}) in the modified $AdS_{5}\times S^{5}$ background
(\ref{globle metric}) which we discussed in the previous section. One
classical solution corresponds to point-like closed strings rotating around
the big circle of $S^{5}$, and the other corresponds to stretched closed
strings rotating around the big circle of $S^{5}$. We then consider the first
order fluctuations around these classical solutions to get the transverse Hamiltonian.

\subsection{Point-like closed string rotating in $S^{5}$}

It is easy to verify that
\begin{align}
t  &  =\nu\tau,\text{ \ \ }\rho=0,\text{ \ \ }\beta_{l}=0\left(
l=1,2,3\right)  ,\text{ \ \ }\nonumber\\
&  \varphi\left(  \equiv\psi_{5}\right)  =\nu\tau,\text{ \ \ }\psi
_{s}=0\left(  s=1,2,3,4\right)  , \label{point-like solution}%
\end{align}
is a solution of the sigma model (\ref{sigma action}) corresponding to the
metric (\ref{globle metric}). This solution describes a point like ($\rho=0$)
closed string boosting around the big circle of $S^{5}$.

To find\ 1-loop approximation, we will consider the fluctuations around the
above classical solution and expand them to the first order. It is useful to
replace $\left(  \rho,\beta_{l}\right)  $ by four Cartesian coordinates
$\eta_{k}\left(  k=1,2,3,4\right)  $ as in \cite{0204226,0209116}
\begin{equation}
\eta_{k}=2r\Omega_{k},\text{ \ \ }\dfrac{2r}{1-r^{2}}=\sinh\rho.
\end{equation}
Written in these Cartesian coordinates, the metric reads
\begin{align}
ds_{B}^{2}  &  =-\left(  \dfrac{1+\eta^{2}/4}{1-\eta^{2}/4}\right)  ^{2}%
dt^{2}+\dfrac{1}{\left(  1-\eta^{2}/4\right)  ^{2}}d\eta_{k}^{2}\nonumber\\
&  \text{ \ \ }+d\psi_{1}^{2}+\cos^{2}\psi_{1}\left(  d\psi_{2}^{2}+\cos
^{2}\psi_{2}d\Omega_{3}^{\prime2}\right) \nonumber\\
&  \text{ \ \ }+\dfrac{4a^{8}x^{i2}u^{8}}{\left(  1+a^{4}u^{4}\right)  ^{2}%
}du^{2}+\dfrac{4a^{4}u^{5}x^{i}}{1+a^{4}u^{4}}dx^{i}du,
\end{align}
and the transformations (\ref{globle}) become
\begin{align}
u  &  =\dfrac{1+\eta^{2}/4}{1-\eta^{2}/4}\cos t-\dfrac{\eta_{4}}{1-\eta^{2}%
/4},\nonumber\\
x^{0}  &  =\dfrac{1+\eta^{2}/4\sin t}{\left(  1+\eta^{2}/4\right)  \cos
t-\eta_{4}},\nonumber\\
x^{\mu}  &  =\dfrac{\eta_{\mu}}{\left(  1+\eta^{2}/4\right)  \cos t-\eta_{4}}.
\end{align}
Now, we are ready to consider the fluctuations around the above classical
solution (\ref{point-like solution}) with large sigma model coupling constant
$\lambda$,
\begin{equation}
t=\nu\tau+\dfrac{1}{\lambda^{1/4}}\tilde{t},\text{ \ \ }\varphi=\nu\tau
+\dfrac{1}{\lambda^{1/4}}\tilde{\varphi},\text{ \ \ }\eta_{k}=\dfrac
{1}{\lambda^{1/4}}\tilde{\eta}_{k},\text{ \ \ }\psi_{s}=\dfrac{1}%
{\lambda^{1/4}}\tilde{\psi}_{s}.
\end{equation}
Up to quadratic fluctuations, the Virasoro constraints of the sigma model can
be written as
\begin{align}
T_{aa}  &  =G_{mn}\partial_{a}X^{m}\partial_{a}X^{n}\nonumber\\
&  =\dfrac{1}{\sqrt{\lambda}}\left[  -2\lambda^{1/4}\nu\partial_{\tau}\left(
\tilde{t}-\tilde{\varphi}\right)  -\partial_{a}\tilde{t}\partial^{a}\tilde
{t}+\partial_{a}\tilde{\varphi}\partial^{a}\tilde{\varphi}-\nu^{2}\left(
\tilde{\eta}^{2}+\tilde{\psi}_{s}^{2}\right)  +\partial_{a}\tilde{\eta}%
_{k}\partial^{a}\tilde{\eta}_{k}+\partial_{a}\tilde{\psi}_{s}\partial
^{a}\tilde{\psi}_{s}\right. \nonumber\\
&  \text{ \ \ \ }\left.  -\dfrac{4a^{4}\sin^{2}\nu\tau\cos^{2}\nu\tau}{\left(
1+a^{4}\cos^{4}\nu\tau\right)  ^{2}}\nu^{2}\tilde{\eta}_{i}^{2}-\dfrac
{4a^{4}\sin\nu\tau\cos^{3}\nu\tau}{1+a^{4}\cos^{4}\nu\tau}\nu\tilde{\eta}%
_{i}\partial_{\tau}\tilde{\eta}_{i}\right]  =0. \label{constrain}%
\end{align}
The energy and angular momentum then are given as
\begin{align}
E  &  =P_{t}=2\int_{0}^{2\pi}\dfrac{d\sigma}{2\pi}\left[  \sqrt{\lambda}%
\nu+\lambda^{1/4}\partial_{\tau}\tilde{t}+\nu\tilde{\eta}^{2}\right.
\nonumber\\
&  \text{ \ \ \ \ \ \ \ \ }+\dfrac{4a^{4}\sin^{2}\nu\tau\cos^{2}\nu\tau
}{\left(  1+a^{4}\cos^{4}\nu\tau\right)  ^{2}}\nu\tilde{\eta}_{i}^{2}%
+\dfrac{2a^{4}\sin\nu\tau\cos^{3}\nu\tau}{1+a^{4}\cos^{4}\nu\tau}\tilde{\eta
}_{i}\partial_{\tau}\tilde{\eta}_{i}\nonumber\\
&  \text{ \ \ \ \ \ \ \ \ \ }\left.  +\frac{4\alpha^{\prime}}{b}\frac
{a^{8}\sin\nu\tau\cos^{5}\nu\tau}{1+a^{4}\cos^{4}\nu\tau}\left(  \tilde{\eta
}^{2}\partial_{\sigma}\tilde{\eta}^{3}-\tilde{\eta}^{3}\partial_{\sigma}%
\tilde{\eta}^{2}\right)  \right] \\
& \nonumber\\
J  &  =P_{\varphi}=2\int_{0}^{2\pi}\dfrac{d\sigma}{2\pi}\left[  \sqrt{\lambda
}\nu+\lambda^{1/4}\partial_{\tau}\tilde{\varphi}-\nu\tilde{\psi}_{s}%
^{2}\right]  ,
\end{align}
and the difference between the energy and angular momentum can be obtained as
\begin{align}
E-J  &  =\int_{0}^{2\pi}\dfrac{d\sigma}{2\pi}\left[  \lambda^{1/4}%
\partial_{\tau}\left(  \tilde{t}-\tilde{\varphi}\right)  +\nu\left(
\tilde{\eta}^{2}+\tilde{\psi}_{s}^{2}\right)  \right. \nonumber\\
&  \text{ \ \ \ }+\dfrac{4a^{4}\sin^{2}\nu\tau\cos^{2}\nu\tau}{\left(
1+a^{4}\cos^{4}\nu\tau\right)  ^{2}}\nu\tilde{\eta}_{i}^{2}+\dfrac{2a^{4}%
\sin\nu\tau\cos^{3}\nu\tau}{1+a^{4}\cos^{4}\nu\tau}\tilde{\eta}_{i}%
\partial_{\tau}\tilde{\eta}_{i}\nonumber\\
&  \text{ \ \ \ }\left.  +\frac{4\alpha^{\prime}}{b}\frac{a^{8}\sin\nu\tau
\cos^{5}\nu\tau}{1+a^{4}\cos^{4}\nu\tau}\left(  \tilde{\eta}^{2}%
\partial_{\sigma}\tilde{\eta}^{3}-\tilde{\eta}^{3}\partial_{\sigma}\tilde
{\eta}^{2}\right)  \right]  . \label{e-j}%
\end{align}
Now we can solve for $\partial_{\tau}\left(  \tilde{t}-\tilde{\varphi}\right)
$ from the constraint (\ref{constrain}), and plug it into the above expression
(\ref{e-j}). After rescaling the fields by $\sqrt{2}\lambda^{1/4}$, we end up
with
\begin{align}
E-J  &  =\dfrac{1}{2\nu}\int_{0}^{2\pi}\dfrac{d\sigma}{2\pi}\left[
-\partial_{a}\tilde{t}\partial^{a}\tilde{t}+\partial_{a}\tilde{\varphi
}\partial^{a}\tilde{\varphi}+\nu^{2}\left(  \tilde{\eta}^{2}+\tilde{\psi}%
_{s}^{2}\right)  +\partial_{a}\tilde{\eta}_{k}\partial^{a}\tilde{\eta}%
_{k}+\partial_{a}\tilde{\psi}_{s}\partial^{a}\tilde{\psi}_{s}\right.
\nonumber\\
&  \text{ \ \ \ }\left.  +\dfrac{4a^{4}\sin^{2}\nu\tau\cos^{2}\nu\tau}{\left(
1+a^{4}\cos^{4}\nu\tau\right)  ^{2}}\nu^{2}\tilde{\eta}_{i}^{2}+\frac
{4\alpha^{\prime}}{b}\frac{a^{8}\sin\nu\tau\cos^{5}\nu\tau}{1+a^{4}\cos^{4}%
\nu\tau}\left(  \tilde{\eta}^{2}\partial_{\sigma}\tilde{\eta}^{3}-\tilde{\eta
}^{3}\partial_{\sigma}\tilde{\eta}^{2}\right)  \right] \nonumber\\
&  \equiv\dfrac{1}{\nu}\int_{0}^{2\pi}\dfrac{d\sigma}{2\pi}\mathcal{H}%
^{\left(  2\right)  }.
\end{align}
Thus $E-J$ is given by the expectation value of the transverse Hamiltonian
\begin{align}
\mathcal{H}^{\left(  2\right)  }  &  =\dfrac{1}{2}\left[  -\partial_{a}%
\tilde{t}\partial^{a}\tilde{t}+\partial_{a}\tilde{\varphi}\partial^{a}%
\tilde{\varphi}+\nu^{2}\left(  \tilde{\eta}^{2}+\tilde{\psi}_{s}^{2}\right)
+\partial_{a}\tilde{\eta}_{k}\partial^{a}\tilde{\eta}_{k}+\partial_{a}%
\tilde{\psi}_{s}\partial^{a}\tilde{\psi}_{s}\right. \nonumber\\
&  \left.  +\dfrac{4a^{4}\sin^{2}\nu\tau\cos^{2}\nu\tau}{\left(  1+a^{4}%
\cos^{4}\nu\tau\right)  ^{2}}\nu^{2}\tilde{\eta}_{i}^{2}+\frac{4\alpha
^{\prime}}{b}\frac{a^{8}\sin\nu\tau\cos^{5}\nu\tau}{1+a^{4}\cos^{4}\nu\tau
}\left(  \tilde{\eta}^{2}\partial_{\sigma}\tilde{\eta}^{3}-\tilde{\eta}%
^{3}\partial_{\sigma}\tilde{\eta}^{2}\right)  \right]  . \label{H2}%
\end{align}
It is clear that there are 8 massive bosonic coordinates with two of them
($\tilde{\eta}_{i},i=2,3$) time dependent. Defining $\tilde{x}^{\pm}=\tilde
{t}\pm\tilde{\varphi}$, the transverse Hamiltonian in equation (\ref{H2}) can
be written in \textquotedblleft light-cone\textquotedblright\-like form
\begin{align}
\mathcal{H}^{\left(  2\right)  }  &  =\dfrac{1}{2}\left[  -\partial_{a}%
\tilde{x}^{+}\partial^{a}\tilde{x}^{-}-\frac{1}{4}\left(  \tilde{\eta}%
^{2}+\tilde{\psi}_{s}^{2}\right)  \partial_{a}\tilde{x}^{+}\partial^{a}%
\tilde{x}^{+}+\partial_{a}\tilde{\eta}_{k}\partial^{a}\tilde{\eta}%
_{k}+\partial_{a}\tilde{\psi}_{s}\partial^{a}\tilde{\psi}_{s}\right.
\nonumber\\
&  \left.  +\dfrac{4a^{4}\sin^{2}\nu\tau\cos^{2}\nu\tau}{\left(  1+a^{4}%
\cos^{4}\nu\tau\right)  ^{2}}\nu^{2}\tilde{\eta}_{i}^{2}+\frac{4\alpha
^{\prime}}{b}\frac{a^{8}\sin\nu\tau\cos^{5}\nu\tau}{1+a^{4}\cos^{4}\nu\tau
}\left(  \tilde{\eta}^{2}\partial_{\sigma}\tilde{\eta}^{3}-\tilde{\eta}%
^{3}\partial_{\sigma}\tilde{\eta}^{2}\right)  \right]  .
\end{align}
A similar Hamiltonian can be also obtained by taking the Penrose limit as in
\cite{0205258}.

\subsection{Closed string rotating in $S^{5}$}

In this section we will consider another classical solution of a closed string
whose center of mass is not moving on $S^{5}$, but spins around that point and
is correspondingly stretched. Again, it is easy to verify that the ansatz
\begin{align}
t  &  =\nu\tau,\text{ \ \ }\rho=0,\text{ \ \ }\beta_{l}=0\left(
l=1,2,3\right) \nonumber\\
\theta\left(  \equiv\psi_{1}\right)   &  =\theta\left(  \sigma\right)  ,\text{
\ \ }\varphi\left(  \equiv\psi_{5}\right)  =\nu\tau,\text{ \ \ }\psi
_{s}=0\left(  s=2,3,4\right)  , \label{solution}%
\end{align}
is a solution for the sigma model (\ref{sigma action}) associated with the
modified $AdS_{5}\times S^{5}$ background (\ref{globle metric}).

The constraint for $\theta$ can be found from the Virasoro constraints
$T_{ab}=0$, which is
\begin{equation}
\left(  \theta^{\prime}\right)  ^{2}=\nu^{2}-\nu^{2}\cos^{2}\theta=\nu^{2}%
\sin^{2}\theta\label{constr}%
\end{equation}

Now we would like to repeat the analysis of the previous section in the case
of the more general solution (\ref{solution}). To determine the quantum string
spectrum to the leading order in the large sigma model coupling constant
$\lambda$, we consider the quantum fluctuation around the classical solution
(\ref{solution})
\begin{align}
t  &  =\nu\tau+\dfrac{1}{\lambda^{1/4}}\tilde{t},\text{ \ \ }\varphi=\nu
\tau+\dfrac{1}{\lambda^{1/4}}\tilde{\varphi},\text{ \ \ }\theta=\theta\left(
\sigma\right)  +\dfrac{1}{\lambda^{1/4}}\tilde{\theta},\nonumber\\
&  \text{ }\eta_{k}=\dfrac{1}{\lambda^{1/4}}\tilde{\eta}_{k},\text{ \ \ }%
\psi_{s}=\dfrac{1}{\lambda^{1/4}}\tilde{\psi}_{s}(s=2,3,4).
\end{align}
Up to quadratic fluctuations, the constraints of the sigma model can be
written as
\begin{align}
T_{aa}  &  =G_{mn}\partial_{a}X^{m}\partial_{a}X^{n}\nonumber\\
&  =\dfrac{1}{\sqrt{\lambda}}\left[  -2\lambda^{1/4}\nu\left(  \partial_{\tau
}\tilde{t}-\cos^{2}\theta\partial_{\tau}\tilde{\varphi}\right)  -\partial
_{a}\tilde{t}\partial^{a}\tilde{t}+\cos^{2}\theta\partial_{a}\tilde{\varphi
}\partial^{a}\tilde{\varphi}\right. \nonumber\\
&  \text{ \ \ \ }+2\lambda^{1/4}\nu\sin\theta\partial_{\sigma}\tilde{\theta
}-2\lambda^{1/4}\nu^{2}\sin\theta\cos\theta\cdot\tilde{\theta}-4\nu\sin
\theta\cos\theta\cdot\tilde{\theta}\partial_{\tau}\tilde{\varphi}\nonumber\\
&  \text{ \ \ \ }-\nu^{2}\left(  \cos^{2}\theta-\sin^{2}\theta\right)
\tilde{\theta}^{2}-\nu^{2}\left(  \tilde{\eta}^{2}+\cos^{2}\theta\cdot
\tilde{\psi}_{s}^{2}\right) \nonumber\\
&  \text{ \ \ \ }+\partial_{a}\tilde{\eta}_{k}\partial^{a}\tilde{\eta}%
_{k}+\partial_{a}\tilde{\theta}\partial^{a}\tilde{\theta}+\cos^{2}%
\theta\partial_{a}\tilde{\psi}_{s}\partial^{a}\tilde{\psi}_{s}\nonumber\\
&  \text{ \ \ \ }\left.  -\dfrac{4a^{4}\sin^{2}\nu\tau\cos^{2}\nu\tau}{\left(
1+a^{4}\cos^{4}\nu\tau\right)  ^{2}}\nu^{2}\tilde{\eta}_{i}^{2}-\dfrac
{4a^{4}\sin\nu\tau\cos^{3}\nu\tau}{1+a^{4}\cos^{4}\nu\tau}\nu\tilde{\eta}%
_{i}\partial_{\tau}\tilde{\eta}_{i}\right] \nonumber\\
&  =0, \label{constrain2}%
\end{align}
where we have used the constraint (\ref{constr}) to eliminate $\theta^{\prime
}$.

The energy and angular momentum are correspondingly
\begin{align}
E  &  =P_{t}=2\int_{0}^{2\pi}\dfrac{d\sigma}{2\pi}\left[  \sqrt{\lambda}%
\nu+\lambda^{1/4}\partial_{\tau}\tilde{t}+\nu\tilde{\eta}^{2}\right.
\nonumber\\
&  \text{ \ \ \ \ \ \ \ \ \ }+\dfrac{4a^{4}\sin^{2}\nu\tau\cos^{2}\nu\tau
}{\left(  1+a^{4}\cos^{4}\nu\tau\right)  ^{2}}\nu\tilde{\eta}_{i}^{2}%
+\dfrac{2a^{4}\sin v\tau\cos^{3}\nu\tau}{1+a^{4}\cos^{4}\nu\tau}\tilde{\eta
}_{i}\partial_{\tau}\tilde{\eta}_{i}\nonumber\\
&  \text{ \ \ \ \ \ \ \ \ \ }\left.  +\frac{4\alpha^{\prime}}{b}\frac
{a^{8}\sin\nu\tau\cos^{5}\nu\tau}{1+a^{4}\cos^{4}\nu\tau}\left(  \tilde{\eta
}^{2}\partial_{\sigma}\tilde{\eta}^{3}-\tilde{\eta}^{3}\partial_{\sigma}%
\tilde{\eta}^{2}\right)  \right]  ,\\
& \nonumber\\
J  &  =P_{\varphi}=2\int_{0}^{2\pi}\dfrac{d\sigma}{2\pi}\left[  \sqrt{\lambda
}\nu\cos^{2}\theta+\lambda^{1/4}\cos^{2}\theta\partial_{\tau}\tilde{\varphi
}-2\lambda^{1/4}\nu\sin\theta\cos\theta\cdot\tilde{\theta}\right. \nonumber\\
&  \text{ \ \ \ \ \ \ \ \ \ }\left.  -2\sin\theta\cos\theta\cdot\tilde{\theta
}\partial_{\tau}\tilde{\varphi}-\nu\left(  \cos^{2}\theta-\sin^{2}%
\theta\right)  \tilde{\theta}^{2}-\nu\cos^{2}\theta\tilde{\psi}_{s}%
^{2}\right]  ,
\end{align}
therefore the difference between energy and angular momentum can be easily
calculated
\begin{align}
E-J  &  =\int_{0}^{2\pi}\dfrac{d\sigma}{2\pi}\left[  \sqrt{\lambda}\nu\sin
^{2}\theta+\lambda^{1/4}\left(  \partial_{\tau}\tilde{t}-\cos^{2}%
\theta\partial_{\tau}\tilde{\varphi}\right)  \right. \nonumber\\
&  \text{ \ \ }+2\lambda^{1/4}\nu\sin\theta\cos\theta\cdot\tilde{\theta}%
+2\sin\theta\cos\theta\tilde{\theta}\partial_{\tau}\tilde{\varphi}\nonumber\\
&  \text{ \ \ }+\nu\left(  \cos^{2}\theta-\sin^{2}\theta\right)  \tilde
{\theta}^{2}+\nu\left(  \tilde{\eta}^{2}+\cos^{2}\theta\tilde{\psi}_{s}%
^{2}\right) \nonumber\\
&  \text{ \ \ }+\dfrac{4a^{4}\sin^{2}\nu\tau\cos^{2}\nu\tau}{\left(
1+a^{4}\cos^{4}\nu\tau\right)  ^{2}}\nu\tilde{\eta}_{i}^{2}+\dfrac{2a^{4}%
\sin\nu\tau\cos^{3}\nu\tau}{1+a^{4}\cos^{4}\nu\tau}\tilde{\eta}_{i}%
\partial_{\tau}\tilde{\eta}_{i}\nonumber\\
&  \text{ \ \ }\left.  +\frac{4\alpha^{\prime}}{b}\frac{a^{8}\sin\nu\tau
\cos^{5}\nu\tau}{1+a^{4}\cos^{4}\nu\tau}\left(  \tilde{\eta}^{2}%
\partial_{\sigma}\tilde{\eta}^{3}-\tilde{\eta}^{3}\partial_{\sigma}\tilde
{\eta}^{2}\right)  \right]  . \label{e-j2}%
\end{align}
As before, we can solve for $\left(  \partial_{\tau}\tilde{t}-\cos^{2}%
\theta\partial_{\tau}\tilde{\varphi}\right)  $ from the constrain
(\ref{constrain2}), and plug it into the above expression (\ref{e-j2}). After
rescaling the fields by $\sqrt{2}\lambda^{1/4}$, we obtain
\begin{align}
E-J  &  =\dfrac{1}{2\nu}\int_{0}^{2\pi}\dfrac{d\sigma}{2\pi}\left[
-\partial_{a}\tilde{t}\partial^{a}\tilde{t}+\cos^{2}\theta\partial_{a}%
\tilde{\varphi}\partial^{a}\tilde{\varphi}+\nu^{2}\left(  \cos^{2}\theta
-\sin^{2}\theta\right)  \tilde{\theta}^{2}+\partial_{a}\tilde{\theta}%
\partial^{a}\tilde{\theta}\right. \nonumber\\
&  \text{ \ \ \ }+\cos^{2}\theta\left(  \nu^{2}\tilde{\psi}_{s}^{2}%
+\partial_{a}\tilde{\psi}_{s}\partial^{a}\tilde{\psi}_{s}\right)  +\nu
^{2}\tilde{\eta}^{2}+\partial_{a}\tilde{\eta}_{k}\partial^{a}\tilde{\eta}%
_{k}\nonumber\\
&  \text{ \ \ \ }+\dfrac{4a^{4}\sin^{2}\nu\tau\cos^{2}\nu\tau}{\left(
1+a^{4}\cos^{4}\nu\tau\right)  ^{2}}\nu^{2}\tilde{\eta}_{i}^{2}+\frac
{4\alpha^{\prime}}{b}\frac{a^{8}\sin\nu\tau\cos^{5}\nu\tau}{1+a^{4}\cos^{4}%
\nu\tau}\left(  \tilde{\eta}^{2}\partial_{\sigma}\tilde{\eta}^{3}-\tilde{\eta
}^{3}\partial_{\sigma}\tilde{\eta}^{2}\right) \nonumber\\
&  \text{ \ \ \ }\left.  +2\nu^{2}\sin\theta\partial_{\sigma}\tilde{\theta
}+2\nu^{2}\sin\theta\cos\theta\cdot\tilde{\theta}+2\nu^{2}\sin^{2}%
\theta\right] \nonumber\\
&  \equiv\dfrac{1}{\nu}\int_{0}^{2\pi}\dfrac{d\sigma}{2\pi}\mathcal{H}%
^{\left(  2\right)  }.
\end{align}
Thus, $E-J$ is given by the expectation value of the transverse Hamiltonian
\begin{align}
\mathcal{H}^{\left(  2\right)  }  &  =-\partial_{a}\tilde{t}\partial^{a}%
\tilde{t}+\cos^{2}\theta\partial_{a}\tilde{\varphi}\partial^{a}\tilde{\varphi
}+\nu^{2}\left(  \cos^{2}\theta-\sin^{2}\theta\right)  \tilde{\theta}%
^{2}+\partial_{a}\tilde{\theta}\partial^{a}\tilde{\theta}\nonumber\\
&  \text{ \ \ }+\cos^{2}\theta\left(  \nu^{2}\tilde{\psi}_{s}^{2}+\partial
_{a}\tilde{\psi}_{s}\partial^{a}\tilde{\psi}_{s}\right)  +\nu^{2}\tilde{\eta
}^{2}+\partial_{a}\tilde{\eta}_{k}\partial^{a}\tilde{\eta}_{k}\nonumber\\
&  \text{ \ \ }+\dfrac{4a^{4}\sin^{2}\nu\tau\cos^{2}\nu\tau}{\left(
1+a^{4}\cos^{4}\nu\tau\right)  ^{2}}\nu^{2}\tilde{\eta}_{i}^{2}+\frac
{4\alpha^{\prime}}{b}\frac{a^{8}\sin\nu\tau\cos^{5}\nu\tau}{1+a^{4}\cos^{4}%
\nu\tau}\left(  \tilde{\eta}^{2}\partial_{\sigma}\tilde{\eta}^{3}-\tilde{\eta
}^{3}\partial_{\sigma}\tilde{\eta}^{2}\right) \nonumber\\
&  \text{ \ \ }+2\nu^{2}\sin\theta\partial_{\sigma}\tilde{\theta}+2\nu^{2}%
\sin\theta\cos\theta\cdot\tilde{\theta}+2\nu^{2}\sin^{2}\theta.
\end{align}
The following transformation
\begin{equation}
\bar{t}=\tilde{t},\text{ \ \ }\bar{\varphi}=\cos\theta\tilde{\varphi},\text{
\ \ }\bar{\psi}_{s}=\cos\theta\tilde{\psi}_{s}\left(  s=2,3,4\right)  ,\text{
\ \ }\bar{\theta}=\tilde{\theta},\text{ \ \ }\bar{\eta}_{k}=\tilde{\eta}%
_{k}\left(  k=0,1,2,3\right)  ,
\end{equation}
will bring the kinetic terms to canonical form. The resulting Hamiltonian is
\begin{align}
\mathcal{H}^{\left(  2\right)  }  &  =-\partial_{a}\bar{t}\partial^{a}\bar
{t}+\partial_{a}\bar{\varphi}\partial^{a}\bar{\varphi}+\partial_{a}\bar
{\theta}\partial^{a}\bar{\theta}+\partial_{a}\bar{\eta}_{k}\partial^{a}%
\bar{\eta}_{k}+\partial_{a}\bar{\psi}_{s}\partial^{a}\bar{\psi}_{s}\nonumber\\
&  \text{ \ \ }+m_{\varphi}^{2}\bar{\varphi}^{2}+m_{\theta}^{2}\bar{\theta
}^{2}+m_{\psi}^{2}\bar{\psi}_{s}^{2}+m_{\lambda}^{2}\bar{\eta}_{\lambda}%
^{2}+m_{i}^{2}\bar{\eta}_{i}^{2}\nonumber\\
&  \text{ \ \ }+\frac{4\alpha^{\prime}}{b}\frac{a^{8}\sin\nu\tau\cos^{5}%
\nu\tau}{1+a^{4}\cos^{4}\nu\tau}\left(  \tilde{\eta}^{2}\partial_{\sigma
}\tilde{\eta}^{3}-\tilde{\eta}^{3}\partial_{\sigma}\tilde{\eta}^{2}\right)
\nonumber\\
&  \text{ \ \ }+\dfrac{2\nu\sin^{2}\theta}{\cos\theta}\left(  \bar{\varphi
}\partial_{\sigma}\bar{\varphi}+\bar{\psi}_{s}\partial_{\sigma}\bar{\psi}%
_{s}\right)  +2\nu^{2}\sin\theta\left(  1+\cos\theta\cdot\bar{\theta}%
+\partial_{\sigma}\bar{\theta}\right)  , \label{H2 - string}%
\end{align}
where
\begin{align}
m_{\varphi}^{2}  &  =\dfrac{\nu^{2}\sin^{4}\theta}{\cos^{2}\theta},\\
m_{\theta}^{2}  &  =\nu^{2}\left(  \cos^{2}\theta-\sin^{2}\theta\right)  ,\\
m_{\psi}^{2}  &  =\nu^{2}\left(  1+\dfrac{\sin^{4}\theta}{\cos^{2}\theta
}\right)  ,\\
m_{\lambda}^{2}  &  =\nu^{2},\lambda=0,1,\\
m_{i}^{2}  &  =\nu^{2}\left(  1+\dfrac{4a^{4}\sin^{2}\nu\tau\cos^{2}\nu\tau
}{\left(  1+a^{4}\cos^{4}\nu\tau\right)  ^{2}}\right)  ,\text{ \ \ }i=2,3.
\end{align}
It is clear that when $\theta\left(  \sigma\right)  =0$, which correspond the
point-like closed string case in the previous section, the Hamiltonian
(\ref{H2 - string}) reduces to the Hamiltonian (\ref{H2}) as expected.

\section{Quantization}

In this section, we will quantize the classical systems we studied in the
previous sections to compute the full closed string spectrum. We see that when
$a=0$, which corresponds to the case of the vanishing background $B$ field,
the Hamiltonians (\ref{H2}) and (\ref{H2 - string}) will reduce to the closed
strings moving in the standard $AdS_{5}\times S^{5}$ background. The
quantization of the closed strings moving in the standard $AdS_{5}\times
S^{5}$ background has been widely studied by many authors, for example in
\cite{0204226,0209116}. To quantize the classical closed string systems in the
modified $AdS_{5}\times S^{5}$ background (\ref{globle metric}), we need to
solve the equations of motion corresponding to the Hamiltonians (\ref{H2}) and
(\ref{H2 - string}) respectively. Since the difference between our current
case and the case of the closed strings moving in the standard $AdS_{5}\times
S^{5}$ background is the $a$ dependent terms, we only need to solve for the
fields whose equations of motion\ are affected by $a$. It is easy to see in
the Hamiltonians (\ref{H2}) and (\ref{H2 - string}) that all the fields have
the exactly same equations of motion as the case without the background $B$
field except for the fields $\tilde{\eta}_{i}$, $i=2,3,$ which are the fields
along the directions with the $B$ field turned on. The Hamiltonians for the
fields $\tilde{\eta}_{i}$, $i=2,3$ coming from (\ref{H2}) and
(\ref{H2 - string}) turn out to be in the same form, which is
\begin{align}
\mathcal{H}_{\tilde{\eta}}  &  =\dfrac{1}{2}\int\dfrac{d\sigma}{2\pi}\left[
\partial_{\tau}\tilde{\eta}_{i}\partial_{\tau}\tilde{\eta}_{i}+\partial
_{\sigma}\tilde{\eta}_{i}\partial_{\sigma}\tilde{\eta}_{i}+\left(
1+\dfrac{4a^{4}\sin^{2}\nu\tau\cos^{2}\nu\tau}{\left(  1+a^{4}\cos^{4}\nu
\tau\right)  ^{2}}\right)  \nu^{2}\tilde{\eta}_{i}^{2}\right. \nonumber\\
&  \text{ \ \ \ }\left.  +\frac{4\alpha^{\prime}}{b}\frac{a^{8}\sin\nu\tau
\cos^{5}\nu\tau}{1+a^{4}\cos^{4}\nu\tau}\left(  \tilde{\eta}^{2}%
\partial_{\sigma}\tilde{\eta}^{3}-\tilde{\eta}^{3}\partial_{\sigma}\tilde
{\eta}^{2}\right)  \right]  . \label{H eta}%
\end{align}
It is hard to solve the equations of motion corresponding to the above
Hamiltonian (\ref{H eta}) because of the complicated time-dependent
coefficient. To proceed, we make an approximation by letting $a\rightarrow0$
but still \emph{finite}. Notice that according to decoupling limit we took in
(\ref{transfer2}) and the definition (\ref{transfer1}) and (\ref{transfer3}),
any \emph{finite} value of $a$, no matter how small it is, corresponds to the
infinite value of the background $B$ field, i.e. $B_{23}\rightarrow\infty$.
This is important for us because only for the infinite background $B$ field
the dual field theory on the boundary is noncommutative field theory. Under
the limit $a\rightarrow0$, we expand the Hamiltonian (\ref{H eta}) in $a$ and
keep the lowest order. The Hamiltonian now is simplified to the form
\begin{equation}
\mathcal{H}_{\tilde{\eta}}=\int\dfrac{d\sigma}{2\pi}\left[  \partial_{\tau
}\tilde{\eta}_{i}\partial_{\tau}\tilde{\eta}_{i}+\partial_{\sigma}\tilde{\eta
}_{i}\partial_{\sigma}\tilde{\eta}_{i}+\left(  1+a^{4}\sin^{2}2\nu\tau\right)
\nu^{2}\tilde{\eta}_{i}^{2}\right]  . \label{HB}%
\end{equation}
It it easy to get the equations of motion for the fields $\tilde{\eta}_{i}$
\begin{equation}
\partial_{\tau}^{2}\tilde{\eta}_{i}+\partial_{\sigma}^{2}\tilde{\eta}%
_{i}+\left(  1+a^{4}\sin^{2}2\nu\tau\right)  \nu\tilde{\eta}_{i}^{2}=0.
\end{equation}
We expand\footnote{We ignore the sub-index $i(=2,3)$ from now for neat
expression.} the field $\tilde{\eta}$ in the different modes as $\tilde{\eta
}=\sum\tilde{\eta}_{n}e^{in\sigma}$, the equation of motion for each mode
$\tilde{\eta}_{n}$ can be obtained as
\begin{equation}
\dfrac{\partial^{2}\tilde{\eta}_{n}}{\partial z^{2}}+\left(  \lambda
_{n}-2q\cos2z\right)  \tilde{\eta}_{n}=0, \label{mathieu}%
\end{equation}
where
\begin{equation}
\lambda_{n}=\dfrac{1}{4\nu^{2}}\left(  n^{2}+\nu^{2}+\dfrac{\nu^{2}}{2}%
a^{4}\right)  ,\text{ \ }q=\dfrac{a^{4}}{16},\text{ \ }z=2\nu\tau.
\end{equation}
Equation (\ref{mathieu}) is a typical Mathieu equation. It is worth noting
that the similar Mathieu equation has been obtained when people studied the
scalar field in the same NS $B$ field background, but here we are considering
the full closed string fields. We refer reader to \cite{9805140,9908134} for
the general method to solve Mathieu equation.

Let us define $Z_{n}^{\pm}$ as the two independent solutions of the equation
(\ref{mathieu}). Then the mode expansion for the field $\tilde{\eta}$ can be
written as
\begin{equation}
\tilde{\eta}=\sum_{n=1}^{\infty}\left[  C_{n}Z_{n}^{-}\left(  \alpha
_{n}e^{in\sigma}+\tilde{\alpha}_{n}e^{-in\sigma}\right)  +C_{-n}Z_{n}%
^{+}\left(  \alpha_{n}^{+}e^{-in\sigma}+\tilde{\alpha}_{n}^{+}e^{in\sigma
}\right)  \right]  , \label{string solution}%
\end{equation}
where $\alpha_{n}/\tilde{\alpha}_{n}$'s and $\alpha_{n}^{+}/\tilde{\alpha}%
_{n}^{+}$'s are annihilation and creation operators, $C_{n}/C_{-n}$'s are
constant coefficients.

To quantize the field $\tilde{\eta}$, we define the conjugate momentum
\begin{equation}
\Pi_{\tilde{\eta}}=\partial_{\tau}\tilde{\eta}=\sum_{n=1}^{\infty}\left[
C_{n}\dot{Z}_{n}^{-}\left(  \alpha_{n}e^{in\sigma}+\tilde{\alpha}%
_{n}e^{-in\sigma}\right)  +C_{-n}\dot{Z}_{n}^{+}\left(  \alpha_{n}%
^{+}e^{-in\sigma}+\tilde{\alpha}_{n}^{+}e^{in\sigma}\right)  \right]  .
\end{equation}
and impose the quantization condition
\begin{equation}
\left[  \tilde{\eta}^{i}\left(  \sigma,\tau\right)  ,\Pi_{\tilde{\eta}}%
^{j}\left(  \sigma^{\prime},\tau\right)  \right]  =i\pi\delta^{ij}%
\delta\left(  \sigma-\sigma^{\prime}\right)  . \label{quantization}%
\end{equation}
The constants $C_{n}$ can be determined using the quantization condition
(\ref{quantization}) as normalization condition
\begin{equation}
C_{n}=C_{-n}=\sqrt{\dfrac{i}{\left(  Z_{n}^{-}\dot{Z}_{n}^{+}-Z_{n}^{+}\dot
{Z}_{n}^{-}\right)  }}, \label{Wronskian}%
\end{equation}
where $\left(  Z_{n}^{-}\dot{Z}_{n}^{+}-Z_{n}^{+}Z_{n}^{-}\right)  $ is the
Wronskian of the Mathieu equation and therefore it is a constant.

The substitution of the solution (\ref{string solution}) into the Hamiltonian
(\ref{HB}) leads to the expression
\begin{align}
\mathcal{H}_{\tilde{\eta}}  &  =\sum_{n=0}^{\infty}C_{n}^{2}\left\{  \left[
\left(  n^{2}+\nu^{2}+\nu^{2}a^{4}\sin^{2}2\nu\tau\right)  Z_{n}^{-}Z_{n}%
^{+}+\dot{Z}_{n}^{-}\dot{Z}_{n}^{+}\right]  \left(  \alpha_{n}^{\dag}%
\alpha_{n}+\tilde{\alpha}_{n}^{\dag}\tilde{\alpha}_{n}\right)  \right.
\nonumber\\
&  \text{ \ \ }+\left[  \left(  n^{2}+\nu^{2}+\nu^{2}a^{4}\sin^{2}2\nu
\tau\right)  Z_{n}^{-}Z_{n}^{-}+\dot{Z}_{n}^{-}\dot{Z}_{n}^{-}\right]
\alpha_{n}\tilde{\alpha}_{n}\nonumber\\
&  \text{ \ \ }+\left.  \left[  \left(  n^{2}+\nu^{2}+\nu^{2}a^{4}\sin^{2}%
2\nu\tau\right)  Z_{n}^{+}Z_{n}^{+}+\dot{Z}_{n}^{+}\dot{Z}_{n}^{+}\right]
\alpha_{n}^{\dag}\tilde{\alpha}_{n}^{\dag}\right\}  .
\end{align}
It is easy to verify that when $a=0$ the last two terms vanish and the above
Hamiltonian reduces to the original one without the background $B$ field.

To understand the Hamiltonian better, we need the explicit expression of the
functions $Z_{n}^{\pm}$. But in general, the solutions of Mathieu equation can
not be written as any known function which is easy to work with. However, in
the limit of $a\rightarrow0$, we can use WKB approximation. In this case the
solutions have a very simple form
\begin{equation}
Z_{n}^{\pm}=\exp\left\{  \pm i\sqrt{\lambda_{n}}\left(  z-\dfrac{q}%
{2\lambda_{n}}\sin2z\right)  \right\}  . \label{WKB approximation}%
\end{equation}
The constant $C_{n}$ can then be determined as\footnote{Directly putting the
solutions (\ref{WKB approximation}) into (\ref{Wronskian}) will produce a
time-dependent $C_{n}$ since (\ref{WKB approximation}) are not exact solutions
of the equation of motion (\ref{mathieu}). We need to use the fact that when
$\tau\rightarrow0$, the solutions should reduce to the ones without background
$B$-field to determine $C_{n}$.}%
\begin{equation}
C_{n}^{2}=\dfrac{\sqrt{n^{2}+\nu^{2}+\dfrac{\nu^{2}a^{4}}{2}}}{2\left(
n^{2}+\nu^{2}+\dfrac{\nu^{2}a^{4}}{4}\right)  }\approx\frac{1}{2\sqrt
{n^{2}+\nu^{2}}}.
\end{equation}
With these functions $Z_{n}^{\pm}$ at hand, the Hamiltonian can be expressed
as
\begin{equation}
\mathcal{H}_{\tilde{\eta}}=2\sum_{n=0}^{\infty}C_{n}^{2}\left(  n^{2}+\nu
^{2}+\nu^{2}a^{4}\sin^{2}2\nu\tau\right)  \left(  \alpha_{n}^{\dag}\alpha
_{n}+\tilde{\alpha}_{n}^{\dag}\tilde{\alpha}_{n}\right)  .
\end{equation}
Finally, the difference between energy and angular momentum $E-J$ is given by
the expectation value of the above Hamiltonian , which is
\begin{align}
E-J  &  =\dfrac{1}{\nu}\sum_{n=0}^{\infty}\left(  \sqrt{n^{2}+\nu^{2}}%
+\dfrac{\nu^{2}a^{4}\sin^{2}2\nu\tau}{2\sqrt{n^{2}+\nu^{2}}}\right)  \left(
N_{n}+\tilde{N}_{n}\right) \nonumber\\
&  =\sum_{n=0}^{\infty}\left(  \sqrt{1+\dfrac{\lambda n^{2}}{J^{2}}}%
+\dfrac{a^{4}\sin^{2}2\nu\tau}{2\sqrt{1+\dfrac{\lambda n^{2}}{J^{2}}}}\right)
\left(  N_{n}+\tilde{N}_{n}\right) \nonumber\\
&  =\sum_{n=0}^{\infty}\left(  \sqrt{1+\dfrac{\lambda n^{2}}{J^{2}}}%
+\dfrac{2a^{4}u^{2}\left\vert 1-u^{2}\right\vert }{\sqrt{1+\dfrac{\lambda
n^{2}}{J^{2}}}}\right)  \left(  N_{n}+\tilde{N}_{n}\right)  ,
\label{quantized e-j}%
\end{align}
where we used $J=\sqrt{\lambda}\nu$ and $u=\cos\nu\tau\in\left[  -1,1\right]
$.

\section{Holographic Noncommutativity}

String field in the modified $AdS_{5}\times S^{5}$ background (\ref{metric})
is conjectured to be dual to the noncommutative Yang-Mills theory on the
boundary \cite{9907166,9908134}. Thus, similar to the analysis in
\cite{0202021}, the string spectrum should correspond one-to-one to certain
operators in the noncommutative Yang-Mills theory. However, the closed string
spectrum (\ref{quantized e-j}), which we found in the previous section, is
time dependent. At first glance it seems strange that the spectrum depends on
time. In fact, this kind of time dependent spectrum has been studied in
\cite{0211289,0205258} when the authors tried to quantize the string fields in
the non-extremal $NS5$-brane background. The crucial point is that the time
$\tau$ in the string spectrum (\ref{quantized e-j}) is not the space-time time
$t$, but the world-sheet time. Actually, the world-sheet time $\tau$ is
related to the space-time $u$ direction, which measures the energy scale in
the holographic description of the boundary field theory
\cite{9912012,0205258}. Therefore, the world-sheet time dependent string
spectrum corresponds to the operators at the different energy scales and can
be interpreted as the RG flow in the dual field theories.%

%TCIMACRO{\FRAME{ftbpFU}{7.7591in}{1.7417in}{0pt}{\Qcb{Closed string
%fluctuation about the classical solution: (a) without the background $B$
%field, the fluctuated string state $a^{\dag}|0\rangle$ is independent of the
%world-sheet time $\tau$; (b) with the background $B$ field, the fluctuated
%string state $a^{\dag}\left(  \tau\right)  |0\rangle$ dependents on the
%world-sheet time.}}{\Qlb{fig1}}{fig1.eps}%
%width 7.7591in;  height 1.7417in;  depth 0pt;  original-width 7.6942in;
%original-height 1.7037in;  cropleft "0";  croptop "1";  cropright "0.9993";
%cropbottom "0";  filename 'fig1.eps';file-properties "XNPEU";}} }%
%BeginExpansion
\begin{figure}
[ptb]
\begin{center}
\includegraphics[
trim=0.000000in 0.000000in 0.005386in 0.000000in,
height=1.7417in,
width=7.7591in
]%
{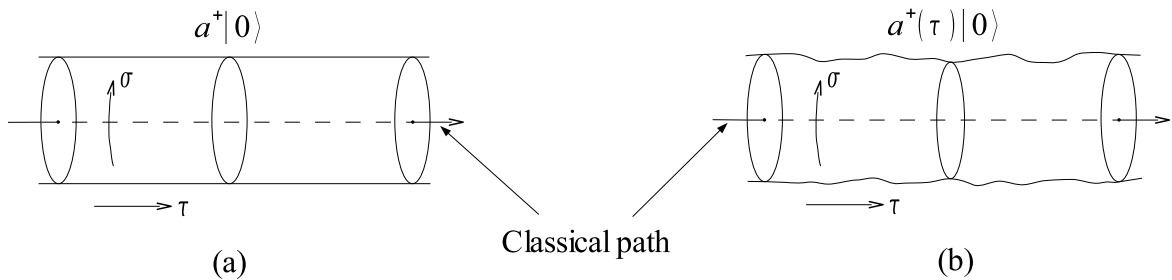}%
\caption{Closed string fluctuation about the classical solution: (a) without
the background $B$ field, the fluctuated string state $a^{\dag}|0\rangle$ is
independent of the world-sheet time $\tau$; (b) with the background $B$ field,
the fluctuated string state $a^{\dag}\left(  \tau\right)  |0\rangle$
dependents on the world-sheet time.}%
\label{fig1}%
\end{center}
\end{figure}
%EndExpansion

When we consider the closed string states created by the string fluctuation
around the classical solution, a string state $a^{\dag}|0\rangle$ corresponds
to a operator $\mathcal{O}$\ in the dual field theory (at certain energy scale
$\Lambda$) \cite{0202021}. In the absence of background $B$ field, the
Hamiltonians of the quadratic fluctuations are time independent, so that the
corresponding string state is the same at different world-sheet time $\tau$
along the classical path, as in (a) of Fig. \ref{fig1}. Thus the string state
corresponds to an operator at the same energy scale in the dual field theory.
In the presence of $B$ field , however, the Hamiltonians of the quadratic
fluctuations (\ref{H2}) or (\ref{H2 - string}) are time dependent. The string
state could absorb or lose energy to change to another string state over time,
so that the string state $a^{\dag}\left(  \tau\right)  |0\rangle$ is different
at different time $\tau$ along the classical path, as indicated in (b) of Fig.
\ref{fig1}. The claim is that a string state varying with the world-sheet time
$a^{\dag}\left(  \tau\right)  |0\rangle$ corresponds to a scale dependent
operator $\mathcal{O}\left(  \Lambda\right)  $ in the dual field theory
running over varying energy scales $\Lambda(\tau)$, i.e. the RG flow in the
dual field theory. The correlation functions of the operators $\mathcal{O}%
\left(  \Lambda\right)  $ should satisfy the standard Callan-Symanzik
equations. In our case, the dual field theory is conjectured to be
noncommutative Yang-Mills theory \cite{9907166,9908134}.

For small $u$ the spectrum (\ref{quantized e-j}) reduces to the string
spectrum without the background $B$ field , which corresponds to the IR regime
of the gauge theory. This is consistent with the expectation that
noncommutative Yang-Mills theory reduces to ordinary Yang-Mills theory at long
distances. The other important regime of the dual gauge theory is the UV
regime. Unfortunately, our result is not able to make any prediction about the
UV regime due to the feature of the string spectrum (\ref{quantized e-j}) that
$u$ has a finite range and the energy oscillates between $u_{\text{min}}=-1$
and $u_{\text{max}}=1$, so that the strings never get to the boundary at
$u\rightarrow\infty$ in our configuration. This is because of the special
classical solution (\ref{point-like solution}) and (\ref{solution}) we used in
our analysis. Both of these solutions only cover the region $\left[
-1,1\right]  $ in the $u$ direction. To study the UV regime of the dual gauge
theory, we need to find other classical path which reaches the boundary of
$AdS$ space at $u\rightarrow\infty$. However, we can read some information
around $u\rightarrow\infty$ directly from the original metric (\ref{metric}).
The coefficient of the terms along the $B$-field directions is invariant under
the transformation $u\rightarrow1/\left(  a^{2}u\right)  $. Thus the physics
at $u\rightarrow\infty$ is equivalent to the physics at $u\rightarrow0$. This
suggests that we can not reach the boundary of $AdS$ at $u\rightarrow\infty
$\ and are not able to define the UV fixed point in the dual noncommutative
Yang-Mills theory. This makes perfect sense since a local field theory is
defined at short distances, and in terms of the $AdS/CFT$ correspondence this
means that the microscopic structure of the theory is encoded on the boundary
of the $AdS$ space; on the other hand, the noncommutative Yang-Mills theory is
a non-local theory with UV/IR mixing, which implies that we should not be able
to define the theory at short distances.

\section{Discussion}

We studied closed strings in the background $D3$-brane with constant $NS$ $B$
field, which is conjectured to be dual to the noncommutative Yang-Mills
theory. The closed string spectrum $E-J$ has been computed in the limit
$a\rightarrow0$, where $a$ is a noncommutativity parameter. When $a=0$,
(\ref{quantized e-j}) reduces to the standard BMN formula corresponding to a
$D3$-brane without the background $B$ field as expected. When $a\neq0$, the
fluctuation string spectrum becomes time dependent and can be interpreted as
the RG flow in the dual noncommutative Yang-Mills theory.

A direct check of this duality would be to compare the closed string spectrum
(\ref{quantized e-j}) with the anomalous dimension of certain operators in the
dual noncommutative Yang-Mills theory. In the case of $\mathcal{N}=4$
Yang-Mills theory, the dual operators are \cite{0202021}%
\begin{equation}
\mathcal{O}\left(  x\right)  =\text{Tr}\left(  Z^{J}\right)  \text{,
\ Tr}\left(  \phi Z^{J}\right)  \ldots\label{dual op}%
\end{equation}
The naive guess of the dual operators in the noncommutative theory would be to
replace the ordinary products in the operators (\ref{dual op}) by the Moyal
star products as%
\begin{equation}
\mathcal{O}_{\text{NC}}\left(  x\right)  =\text{Tr}\left(  Z^{\ast J}\right)
\text{, \ Tr}\left(  \phi^{r}\ast Z^{\ast J}\right)  \ldots\label{NC dual op}%
\end{equation}
where%
\begin{equation}
\text{Tr}Z^{\ast J}\equiv\text{Tr}\underset{J}{\underbrace{\left(  Z\ast
Z\ast\cdots\ast Z\right)  }},
\end{equation}
and $Z=\phi^{5}+i\phi^{6}$, $\phi^{r}$, $r=1,2,3,4$ are 6 transverse scalars
in the field theory on the $D$ brane.

However, the operators in (\ref{NC dual op}) are not gauge invariant. To
construct gauge-invariant operators in coordinate space, we define
\cite{0008075}%
\begin{equation}
\mathcal{\hat{O}}_{\text{NC}}\left(  k\right)  =\int d^{4}x\mathcal{O}%
_{\text{NC}}\left(  x\right)  \ast W\left(  x,C\right)  \ast e^{ik\cdot x},
\label{op k}%
\end{equation}
where%
\begin{equation}
W(x,C)=P_{\ast}\exp\left(  ig\int_{C}d\sigma\frac{d\zeta^{\mu}}{d\sigma}%
A_{\mu}\left(  x+\zeta\left(  \sigma\right)  \right)  \right)  .
\end{equation}
Such an operator will be gauge invariant.

Gauge invariant operators in coordinate space can also be defined by a
noncommutative Fourier transformation as\footnote{However, the above procedure
of defining gauge invariant operators in coordinate space is not so clear
since these operators require momentum dependent regularization in
perturbation theory \cite{0008075}.} \cite{0008144,0102158,0201161}
\begin{equation}
\mathcal{\hat{O}}_{\text{NC}}\left(  \hat{x}\right)  =\int d^{4}%
y\mathcal{O}_{\text{NC}}\left(  y\right)  \delta^{\left(  4\right)  }\left(
\hat{x}-y\right)  , \label{op x}%
\end{equation}
where the noncommutative $\delta$-function%
\begin{equation}
\delta^{\left(  4\right)  }\left(  \hat{x}-y\right)  \equiv\int\frac{d^{4}%
k}{\left(  2\pi\right)  ^{4}}e^{-ik\cdot y}e^{ik\cdot\hat{x}}=\int\frac
{d^{4}k}{\left(  2\pi\right)  ^{4}}e^{-ik\cdot y}W\left(  x,C\right)  \ast
e^{ik\cdot x}.
\end{equation}
The next job is to calculate the correlation function of the operators
$\mathcal{\hat{O}}_{\text{NC}}$ in (\ref{op k}) or (\ref{op x}) to get their
anomalous dimensions, and compare them to the string spectrum
(\ref{quantized e-j}). We will postpone this to future work.

\begin{acknowledgments}
Y.Y. would like thank H. Dorn, K. Furuta, Y. Kitazawa, F. Lin, Y. Sekino and
C. Zhu for discussions, and also thank the Physics division of National Center
for Theoretical Sciences (NCTS) of Taiwan and High Energy Accelerators
Research Organization (KEK) of Japan for kind hospitality. This work was
supported in part by an operating grant from the Natural Sciences and
Engineering Research Council of Canada. Y.Y. is also partially supported by a
postdoc fellowship from the National Science Council of Taiwan.
\end{acknowledgments}

\end{document}